\documentclass[12pt]{article}

\usepackage[a4paper, total={6in, 9in}]{geometry}
\usepackage{amssymb}
\usepackage{amsmath}
\usepackage{enumitem}
\usepackage{amsfonts}
\usepackage{bm}

\begin{document}

\centerline{\Large A Characterization of Turing Machines}\smallskip

\centerline{\Large  that Compute Primitive Recursive Functions}  \bigskip

\centerline{Daniel G. Schwartz}

\centerline{Florida State University}

\centerline{Department of Computer Science}

\centerline{Tallahassee, Florida, 32306}

\centerline{\tt dgschwartz@fsu.edu} \bigskip

\begin{abstract}
This paper provides a new and more direct proof of the assertion that a Turing computable function of the natural numbers is primitive recursive if and only if the time complexity of the corresponding Turing machine is bounded by a primitive recursive function of the function's arguments. In addition, it provides detailed proofs of two consequences of this fact, which, although well-known in some circles, do not seem to have ever been published.  The first is that the Satisfiability Problem, properly construed as a function of natural numbers, is primitive recursive.  The second is a generalization asserting that all the problems in NP are similarly primitive recursive.   The purpose here is to present these theorems, fully detailed, in an archival journal, thereby giving them a status of permanence and general availability.  
\end{abstract}

\section{Introduction}

It is sometimes asserted as common knowledge that any Turing computable function of the natural numbers whose Turing machine has time complexity that is bounded by a primitive recursive function of the arguments must itself be primitive recursive.  This typically is a quotation from the 1978 textbook by Machtey and Young \cite{machtey:1978}. That book does provide a proof of this statements, but it has the drawback that the result regarding Turing machines is buried in a somewhat lengthy discussion involving the equivalence of four models of computation, namely, partial recursive functions, random access machines, Markov algorithms, and Turing machines.  In consequence, the intuition concerning the latter becomes lost amidst this complexity.  In addition, a problem arises that the referenced textbook is now long out of print and difficult to obtain.  This is what has motivated my submitting the current work for publication.

The relevant results in \cite{machtey:1978} are Propositions 1.10.1, 1.10.2, and 1.10.4, pp. 54--55.  These read as follows.

\begin{quote}
1.10.1 PROPOSITION \ \ Suppose a function $f$ is computed by a program whose space and time complexity is bounded by a primitive recursive or elementary function in any one of the RAM, Turing machine, Markov algorithm, or partial recursive programming systems.  Then $f$ has programs which compute it in space and time bounded by a primitive or recursive or elementary function, respectively, in all of these programming systems.
\end{quote}

\begin{quote}
1.10.2 PROPOSITION \ \ Let $f$ be a primitive recursive or elementary function; then $f$ has programs which compute it in time and space bounded by a primitive recursive or  elementary function, respectively, in all of the programming systems mentioned in the previous proposition.
\end{quote}

\begin {quote}
1.10.4 PROPOSITION\ \ Suppose a function $f$ is computed by a program in space or time bounded by a primitive recursive or elementary function in any of the programming systems mentioned in Proposition 1.10.1.  Then $f$ is primitive recursive or elementary respectively. 
\end {quote} 

Most of the effort in \cite{machtey:1978} entails establishing Proposition 1.10.1.  Precise measures of time and space complexity are provided for each of the four programming systems.  Then these systems are related to one another in terms of their complexity measures, first between random access machines (RAM) and Turing machines (TM), then between TM and Markov algorithms (MA), and finally between MA and partial recursive functions (PRF).   Propositions 1.10.2 and 1.10.4 follow from this, with the crucial steps in the latter applying some results regarding Markov algorithms from a preceding chapter.   

The objective in this paper is to provide a simpler and more direct proof of the main result regarding Turing machines and to publish this in an archival journal where this will be more readily available to a broader audience. 

\section{The Main Result}

The proof presented in  this paper is based on the well-known work by Hans Hermes \cite{hermes:1965}.  That book shows how, given a Turing machine $M$ that computes some function $F$ of natural numbers, one can construct a $\mu$-recursive function $G$ that simulates $M$ so that, in effect, $G=F$.  The construction of $G$ employs the well-known Kleene normal form \cite{kleene:1952}, which uses a single application of the $\mu$ operator.  The proof that follows here simply unravels the details, showing that, if the time complexity of $M$ is bounded by a primitive recursive function of the arguments for $M$ (and $F$), then this $\mu$-operator can be replaced with a bounded $\mu$-operator, so that the Kleene normal form, i.e., $G$, becomes primitive recursive.  

\subsection{Primitive Recursive Functions}

There are various equivalent formulations of primitive recursion.  Since this work relies on the work by Hermes, it follows that treatment closely, with only minor differences in the choice of notations.  These differences will be noted wherever they occur.  This treatment additionally provides items not found in \cite{hermes:1965} that are needed for present purpose.  

The {\it primitive recursive functions} are functions that take natural numbers as arguments and produce natural numbers as values.  They are defined by providing a collection of {\it initial functions} and some schemas for definition by {\it substitution} and {\it recursion} and then stating that the primitive recursive functions comprise all those functions that can be generated from the initial functions by repeated applications of these two schemas.  The definition given here is adapted from Hermes \cite{hermes:1965}, pp. 60-61, but also borrows from Davis and Weyuker \cite{davis:1983}.

For each natural number $x$, let $x'$ denote the {\it successor of} $x$, i.e., the number $x+1$.  The {\it primitive recursive functions} may be defined as follows.

\begin{enumerate}

\item The initial functions: 

\begin{enumerate}
\item the {\it successor function} $S$ defined by $S(x)=x'$, 

\item the {\it constant zero function} $Z$ defined by $Z(x)=0$, for all $x$, and, 

\item for each $i$ and $n$, the {\it n-ary projection function} $P^n_i$ defined by \hfill\break $P^n_i(x_1,\ldots,x_n)=x_i$,

\end{enumerate}

are primitive recursive.

\item If  $G$ is an $m$-ary primitive recursive function, $m\ge 1$, and $H_1,\ldots,H_m$ are $n$-ary primitive recursive functions, $n\ge 1$, then the {\it composition} $C^n_m[G,H_1,\ldots,H_m]$ defined by \medskip

$C^n_m[G,H_1,\ldots,H_m](x_1,\ldots,x_n)=G(H_1(x_1,\ldots,x_n),\ldots,H_m(x_1,\ldots,x_n))$ \hfill(*) 

\medskip

is primitive recursive.

\item For $n\ge 1$, if $G$ is $n$-ary, $H$ is $n+2$-ary, and both are primitive recursive, then the {\it recursion} $R^n[G,H]$ defined by \medskip

\setlength{\tabcolsep}{2pt}
$
\begin{tabular}{l c l}
$R^n[G,H](x_1,\ldots,x_n,0)$&$=$&$G(x_1,\ldots,x_n)$ \\
$R^n[G,H](x_1,\ldots,x_n,x')$&$=$&$H(x_1,\ldots,x_n,x,R^n[G,H](x_1,\ldots,x_n,x))$
\end{tabular}
\Bigg\}$ \hfill (**) \medskip

is primitive recursive.

\item No functions are primitive recursive except as required by items 1--3.
\end{enumerate}

\noindent A difference between \cite{hermes:1965} and \cite{davis:1983} is that the former does not include the zero function among the initial functions, but instead allows the $G$ in the composition schema to be 0-ary and then takes the constant 0-ary function $C^0_0$, which evaluates to 0, as an initial function, whereas the latter does what is used here.   That these two approaches are equivalent is in fact explicitly stated by Hermes on p. 37: ``We shall denote the function of zero arguments which has the value $W$ by $C^W_0$, or also in short by $W$ if no misunderstanding may result from it."  Thus we are here taking $W$ as 0 and using this in place of $C^0_0$.  The use of the  function $Z$ also borrows from  \cite{davis:1983}, although their notation for this is $n$. 

In \cite{hermes:1965}, the projection functions are called ``identity functions", and the letter $U$ is used.  It is not clear to me why this letter was chosen, but I do note that \cite{davis:1983} seems to have followed suit by using the same terminology and the lower case $u$.  However, I have more recently noted that others have called these {\it projection functions} and use the letter $P$ (e.g., \cite{wikipedia:2024}).  In retrospect, this seems to be more appropriate, so I have adopted this here.   

The composition, $C^n_m[G,H_1,\ldots,H_m]$, is also referred to as the {\it substitution} of $H_1,\ldots,H_m$ {\it in} $G$; the schema (*) is referred to as the {\it composition schema} or the {\it substitution schema}; and functions defined using this schema are said to be defined {\it by composition} or {\it by substitution}.  The recursion, $R^n[G,H]$, is also referred to an instance of {\it induction}; the schema (**) is referred to as the {\it recursion schema} or the {\it induction schema}, and functions defined using this schema are said to be defined {\it by recursion} or {\it induction}.

So far, this presentation has mostly repeated material from Hermes \cite{hermes:1965}, with only slight notational changes.  The following assumes that the reader is familiar with Sections 10 and 11 of \cite{hermes:1965}, which cover primitive recursive functions and predicates in detail.  

\subsection {The Characterization Result}

We will make use of Hermes' proof of the equivalence of the Turing machines that compute functions of the natural numbers with the $\mu$-recursive functions \cite{hermes:1965}.  Chapter 2 of that work presents a detailed definition of a Turing machine over an alphabet $\mathfrak{A}=\{a_1,\ldots,a_N\}$ and defines what it means for a $k$-ary function of words over $\mathfrak{A}$ to be {\it Turing-computable}.  Chapter 3 introduces the $\mu$-recursive functions, these being the functions of natural numbers that are obtained from the primitive recursive functions by adding the unbounded $\mu$-operator (see Section 12, pp. 74--75), which can be applied to predicates to turn them into functions.  For example, if $P$ is a $(n+1)$-ary predicate, one may define a $n$-ary function $f$ by \medskip

$f(x_1,\ldots,x_n)=\mu yP(x_1,\ldots,x_n,y)$ \medskip
 
\noindent which evaluates as the smallest $y$ such that $P(x_1,\ldots,x_n,y)$ holds, if there is an $(n+1)$-tuple for which $P$ holds, and evaluates to 0 if there is not. 

Section 15, p. 94, states the two main theorems: \medskip

{\it Theoren A. Every $\mu$-recursive function is Turing-computable.} \medskip

{\it Theorem B. Every Turing-computable function is $\mu$-recursive.} \medskip 

\noindent Here is it important to note the statement on that page: 

\begin{quote}

``These theorems show that the class of $\mu$-recursive functions coincides with that of Turing-computable functions.  In showing this we shall always assume, in this and the following two paragraphs, that the {\it the functions mentioned are defined for all $n$-tuples of natural numbers and that the values are natural numbers.}"

\end{quote}

\noindent Thus, more exactly, Theorem B is interpreted as stating that every Turing-computable function {\it of natural numbers} is $\mu$-recursive, and this interpretation should be carried forward in the succeeding Sections 16, 17 and 18.  In particular, in Section 18, paragraph 2, where Hermes defines the function $K(t,\mathfrak{x},z)$, and states

\begin{quote}

``We start from an {\it arbitrary} Turing machine {\bf M} with the G\"odel number $t$. Further let $\mathfrak{x}$ be an arbirary $n$-tuple of arguments."

\end{quote}   

\noindent where $\mathfrak{x}$ denotes the $n$-tuple $\{x_1,\ldots,x_n\}$,  this more exactly means an arbitrary Turing machine {\it that computes a function of natural numbers} and an arbitrary $n$-tuple of {\it natural number} arguments.   In other words, we are here talking about Turing machines whose alphabets consist of only one symbol $a_1$, typically denoted by the stroke, $|$, and not Turing machines over an arbitrary alphabet $\mathfrak{A}=\{a_1,\ldots,a_N\}$, where $N>1$, as might be implied by earlier discussions.  

Continuing with the development in Hermes, Section 15, p. 97, we shall be concerned only with Theorem B and its proof in Sections 17 and 18 in the following modified form: \medskip 

{\it Theorem B$_0$}. There exists a unary primitive recursive function $U$ and, for each $n$, an $(n+2)$-ary primitive recursive predicate $T_n$, with the property that for every $n$-ary Turing-computable function $f$ there exists a number $t$ such that 

\begin{enumerate} 

\item for every sequence of numbers $x_1,\ldots,x_n$, there exists a $y$ with $T_n(t,x_1,\ldots,x_n,y)$, and

\item $f(x_1,\ldots,x_n)=U(\mu y T_n(t,x_1,\ldots,x_n,y))$ for every sequence of numbers \hfill\break $x_1,\ldots,x_n$.

\end{enumerate}  

\noindent Here again note that by ``function $f$" is meant ``function $f$ {\it of natural numbers}".  Theorem B$_0$ confirms that if a given function $f$ of natural numbers is Turing computable, then it is also $\mu$-recursive and may defined in the manner stated.  

This is all that we need to know.  The full proof is given in Section 18, with the foregoing definition for $f(x_1,\ldots,x_n)$ restated on p. 97.  $T_n$ is the well-known predicate of S.C. Kleene's Normal Form Theorem \cite{kleene:1952}, which Hermes also states and proves on p. 97.  

The question, then, is what additional conditions will ensure that the $f$ thus defined is primitive recursive.  Section 12, p. 75, of \cite{hermes:1965} defines the {\it bounded $\mu$-operator}: \medskip

$\mu_{y=0}^k P(x_1,\ldots,x_n,y) = \left\{\begin{array}{l}
\mbox{the smallest $y$ between 0 and $k$ (inclusive) for which} \\
\mbox{\qquad \ $P(x_1,\ldots,x_n,y)$ holds, if such a $y$ exists,} \\
\mbox{0, if no such $y$ exists.}
\end{array}
\right.$ \medskip

\noindent The following is proven on pp. 75--76: \medskip

\begin{quote}
{\it Theorem.} Let $P$ be a primitive recursive predicate.  Let \medskip

$f(x_1,\ldots,x_n,y)=\mu^k_{z=0}P(x_1,\ldots,x_n,y)$. \medskip

\noindent Then $f$ is a primitive recursive function. \medskip
\end{quote}

This theorem can be extended in the following manner.  In the foregoing definition of the bounded $\mu$-operator, replace the two occurrences of $k$ with occurrences of $b(x_1,\ldots,x_n)$, with the stipulation that $b$ is an $n$-ary function. \medskip

\begin{quote}
{\it Corollary.} Let $P$ be a primitive recursive predicate, and let $B$ be a primitive recursive function.  Let \medskip

$f(x_1,\ldots,x_n,y)=\mu^{B(x_1,\ldots,x_n)}_{z=0}P(x_1,\ldots,x_n,y)$. \medskip

\noindent Then $f$ is a primitive recursive function. \medskip
\end{quote}

\noindent This can be established as follows.  Define \medskip

$h(x_1,\ldots,x_n,y,k)=\mu^k_{z=0}P(x_1,\ldots,x_n,y)$. \medskip

\noindent Function $h$ is primitive recursive by the theorem.  Apply the primitive recursion composition (substitution) schema to define \medskip

$f(x_1,\ldots,x_n,y) = h(x_1,\ldots,x_n, y, B(x_1,\ldots,x_n))$. \medskip

\noindent This shows that $f$ is primitive recursive. \medskip

It follows that, if the $\mu$-operator in Theorem B$_0$ can be replaced with a bounded $\mu$-operator that employs a primitive recursive bound, then the defined function $f$ will be primitive recursive.  

In order to determine whether such a bounding function exists, it is necessary look into the details of Hermes' concept of Turing machine, his G\"odel numbering of the relevant data structures, and the functions used to represent the machine's execution.  Here follows a brief review of the relevant details.

The definition of Turing machine and the concept of a computation on such a machine is given in Chapter 2, Sections 5 and 6. The method for G\"odel numbering these machines is presented in Chapter 4, Section 17.  In this section it is assumed that the Turing machine works with a finite alphabet of symbols $\mathfrak{A}=\{a_1,\ldots,a_N\}$, $N\ge 1$, and  a finite set of states $\{c_1,\ldots,c_M\}$.  The notation $a_0$ represents the empty symbol (blank space).  This type of machine uses a tape of cells that is infinite in both directions.  Some cell is indexed by 0; all cells to the right of this are indexed by even numbers; and all cells to the left are indexed by odd numbers.  

A {\it tape expression} is a finite set of words of finite length over the alphabet $\mathfrak A$, with successive words separated by a single occurrence of $a_0$, written onto the tape starting in the cell with index 2 (i.e., immediately to the right of cell 0, as explained in Section 18.2, p. 108), one symbol per cell, and having the remainder of the tape in both directions filled with occurrences of $a_0$.  For each tape expression there is assumed to be a mapping $\beta$ such that, if $j$ is the index of a tape cell, $\beta(j)$ is the index of the symbol on the tape in that cell.  Then, each tape expression is represented by a number $b$ computed as \medskip

$b= \prod_{j=0}^\infty p_j^{\beta(j)}$ \medskip

\noindent where $p_0, p_1, p_2 \ldots$ are the prime numbers $2, 3, 5, \ldots$. Note that for cells $j$ containing the blank space, $\beta(j)=0$, so that all primes raised to this power yield the value 1 and do not contribute to the value of $b$.  Accordingly, for each tape expression, the corresponding $b$ will be finite.  For brevity, this is referred to a ``the tape expression'' $b$, where this is understood to be the encoding of a tape expression.  The concept of a {\it configuration} for a Turing machine is defined in Chapter 2, pp. 23-33.  This is a triple $(a,b,c)$, where $a$ is the index of a cell (implicitly indicating the position of the read/write head), $b$ is a tape expression, and $c$ is the index of a state.  The Turing machine begins its execution with an {\it initial configuration}.  Here the tape expression is the set of arguments being input to the Turing machine, positioned on the tape so that it starts immediately after the cell with index 0 (as explained in Section 18.2, p. 108).  Then each step in the execution of the machine yields a new configuration, thus generating a sequence of configurations, eventually ending in a {\it terminal configuration}, if indeed the Turing machine eventually terminates, and otherwise generating an infinite sequence of such configurations.  A function $K$ is defined (in Section 18.2, pp. 108-110), such that $K(t,x_1,\ldots,x_n,z)$ is the G\"odel number of the $z$-th configuration in this sequence. 

It is useful here to note the $\sigma$-functions defined on pp. 77--78, which assign G\"odel numbers to pairs and triples: \medskip

$\sigma_2(x,y)=2^x(2y +1) \dot - 1$, \medskip

$\sigma_3(x,y,z)=\sigma_2(\sigma(x,y),z)$, \medskip

\noindent as well as the associated functions that extract the elements of the pairs and triples from these numbers.

In Section 18, pp. 108--110, a configuration $(a,b,c)$ is encoded as $\sigma_3(a,b,c)$, and a function $K$ is defined such that $K(t,x_1,\ldots,x_n,z)$ is this encoding of the $z$-th configuration in the sequence of configurations that is generated by applying the Turing machine with G\"odel number $t$ to the arguments $x_1,\ldots,x_n$. 

Analysis of the predicate $T_n$ shows that the smallest $y$ returned by $\mu y$ will be the number of a pair, say $(r,s)$, where $r$ is the index of the terminal configuration in abovementioned configuration sequence, and $s$ is the number $K(t,x_1,\ldots,x_n,s)$, i.e., the encoding of the terminal configuration.  Thus, in order that $y$ be bounded by a primitive recursive function, it is necessary that both $r$ and $s$ be bounded by primitive recursive functions.  Note that a bound on the total number of steps required by the Turing machine's computation would be a bound on $r$, since this is what $r$ represents.  To determine a bound on $s$, it is necessary to look at the components $a,b,c$ of the terminal configuration and determine whether these have primitive recursive bounds.  First, $a$ is the index of the cell where the Turing machine halts.  Since the Turing machine changes position by either 0 or 1 cells on each execution step, this index is certainly bounded by the total number of execution steps, i.e., $r$.   So a bound on $r$ is a bound on $a$.  Second, $b$ is the terminal tape expression, or more exactly, an encoding of a tape expression as defined in the foregoing.  Since the Turing machine writes at most one new symbol on the tape with each execution step, the total number of symbols on the tape is bounded by the number $n$ of initial arguments plus $r$ the number of execution steps.  Clearly, if $r$ has a primitive recursive bound, then so does $n+r$.  This in turn implies that there is a primitive recursive bound on the tape expression encoding $b$.  Last $c\le M$, the total number of states in the Turing machine with G\"odel number $t$, so $M$ is a primitive recursive bound on $c$.

This shows what may be stated here as \medskip

{\bf Proposition 1.} There is a primitive recursive bound on  $\mu y$ in the foregoing definition of $f$ if only there is a primitive recursive bound on the total number of steps in the execution of the Turing machine having G\"odel number $t$ on the arguments $x_1,\ldots,x_n$, i.e., an n-ary primitive recursive function $B$, such that, for each choice of $x_1,\ldots,x_n$, $B(x_1,\ldots,x_n)$ is $\ge$ the total number steps in the execution of indicated Turing machine.  \medskip

This yields the desired result. \medskip

{\bf Theorem 1.} Let {\bf M} be a Turing machine that computes a function $F$ of natural numbers.  If the time complexity of {\bf M} is bounded by a primitive recursive function of the same arguments as $F$, then $F$ is primitive recursive.  \medskip  

While this is the most frequently cited result, the converse proposition is also true.  This can be stated as: \medskip

{\bf Theorem 2.}   Let {\bf M} be a Turing machine that computes a function $F$ of natural numbers.  If $F$ is primitive recursive., then the time complexity of {\bf M} is bounded by a primitive recursive function of the same arguments as $F$.  \medskip  

This may be established by induction on the definition of the primitive recursive functions.  Let us use $\bar x$ to denote the sequence $x_1\ldots,x_n$, and $\tau_F(\bar x)$ to denote the the number of steps taken by the Turing machine {\bf M} to complete the computation of $F(\bar x)$.  Take  $\tau_F(\bar x)$ to be the {\it time complexity} of {\bf M}.  We first show that the time complexities of the elementary functions are bounded by primitive recursive functions.   This will be based on  the presentation of these functions in \cite{hermes:1965}, p. 98.

For the purposes of Turing machine computations, the natural number $n$ is represented by a {\it word} consisting of $n+1$ vertical strokes, $\bm |$.  A Turing machine is assumed to start its computation with the read-write head immediately to the right of set of words representing the arguments to which it is being applied. 

A Turing machine that computes the successor function $S$ is given as ${\bf K{\bm |}r}$, where {\bf K} is the ``copying machine" defined on p. 53.  The action of ${\bf K{\bm |}r}$ is to copy the word to which it is applied, add one stroke to the right, and then move the read-write head one further step to the right.  From  the given definition of the copying machine, is is clear that the time complexity of {\bf K} is a linear function of the numerical value of the word to which is applied.  It follows that the time complexity of ${\bf K{\bm |}r}$ is a linear function $F$ of the value of the word to which it is applied.  Linear functions are primitive recursive, so $F$ is a primitive recursive bound on the time complexity $\tau_S(x)$, of the Turing machine that computes $S(x)$, for all $x$.

The constant zero function $Z$ is computed by the Turing machine $\bf r{\bm |}r$, which moves the read-write head one space to the right, writes a $\bm |$ (representing the number zero), and moves the read-write head one further space to the right.  Thus the time complexity of $Z$ is $\tau_Z(x) = 3$, for all $x$.

A Turing machine that computes the $n$-ary projection function, $P_i^n$, is given as ${\bf K}_{n+1-i}$, where ${\bf K}_m$ is the $m$-copying machine discussed on p. 54 (with $m$ here in place of $n$).  Hermes does not provide details for this machine, but only says the following:

\begin{quote}
We often have the task (especially when computing functions of several arguments) of copying a word which is not placed on the very right, so that the copying procedure has to be carried out over a few words printed in between.   The copying machine ${\bf K}_n$ ($n\ge 1$) carries out this computation.  {\it We assume for this computation that the words in question are separated by gaps of one square only.}  ${\bf K}_1$ is identical to {\bf K}.  ${\bf K}_n$ is built according to  the pattern of {\bf K} with the difference that the uninteresting words lying in between are jumped over every time.  We should bear in mind that $n$ is a fixed number; for every $n$ there exists a machine ${\bf K}_n$.
\end{quote}

This suggests that, if the Turing machine ${\bf K}_n$ is started in a position  immediately to the right of a sequence of words, it will copy the $n$-th word to the left of this starting position.  For the projection function, $P_i^n$, one has a sequence of $n$ words, and the word to be copied is the one that is $n+1-i$ to the left.  This explains the given definition.   Although, the details are omitted, it seems clear that the time complexity of this operation will be a linear function of the sum of the lengths of the words being skipped and the length of the word being copied.  Again, since linear functions are primitve recursive, this provides a primitive recursive bound on the time complexity of this operation.      

Suppose that  $F$ is defined as a composition $C^n_m[G,H_1,\ldots,H_m]$, where the time complexities of $G$ and the $H_i$ are given as primitive recursive functions $\tau_G, \tau_{H_i}, \ldots,\tau_{H_m}$.  Then,  it is easy to see that \medskip

$\tau_F(\bar x)=\tau_G(H_1(\bar x),\ldots,H_m(\bar x))+\sum_{1\le i\le m}\tau_{H_i}(\bar x)$.\medskip

\noindent This is clearly primitive recursive.  

Last suppose that $F$ is defined as a recursion $R^n[G,H]$, where the time complexities of $G$ and $H$ are given as primitive recursive functions $\tau_G$ and $\tau_H$.  Then we have \medskip

$\tau_F(\bar x, x) =\tau_G(\bar x)+\sum_{y<x}\tau_H(\bar x,y,F(\bar x ,y))$. \medskip

\noindent This also is primitive recursive. Thus we have the desired primitive recursive bounds on the time complexities of the relevant Turing machines.

\section{The Satisfiability Problem}

The {\it satisfiability problem} is that of determining whether a Boolean expression is satisfiable.  This can be made precise as follows.  The language of Boolean expressions employs as {\it symbols} the {\it logical connectives} $\lnot$, $\lor$, $\land$, the {\it parentheses} ( and ), and some {\it expression variables} ${\bf e}_1,{\bf e}_2,\ldots$.  The {\it Boolean expressions} are defined by: 

\begin{enumerate}
\item expression variables are Boolean expressions,
\item if $E$ is a Boolean expression, then so is $\lnot E$,
\item if $E$ and $F$ are Boolean expressions, then so are $(E\lor F)$ and $(E\land F)$,
\item nothing is a Boolean expression except as defined by items 1--3. 
\end{enumerate} 

A Boolean expression is {\it satisfiable} if its expression variables can be assigned the values {\bf T} or {\bf F} in such a way that the expression evaluates to {\bf T} using the usual interpretations of the logical connectives.  The satisfiability problem is often stated as a decision problem for a language: \medskip

${\it SAT} =\{E| E {\rm \ is\ a\ satisfiable\ Boolean\ expression}\}$,  \medskip

\noindent i.e., given a Boolean expression $E$, determine whether it is in SAT.

The well-known ``brute-force'' method for deciding {\it SAT} is, given a Boolean expression $E$, construct the truth table for $E$, and determine if the last column contains a {\bf T}.  If so, the answer is ``Yes''; otherwise, the answer is ``No''.  It is desired to show that the function that carries out this computation is primitive recursive.  

First of all, prima facie it does not make sense to ask whether such a function should be primitive recursive, since primitive recursive functions are mappings defined on the natural numbers, and Boolean expressions are strings of symbols.  However, there is a straightforward way to construe this decision problem as a function of numbers.  This is to provide the Boolean expressions with a G\"odel numbering and consider a function, $S$, that takes such a number $x$ as its argument and evaluates to 1 if $x$ is the G\"odel number of a satisfiable Boolean expression, and evaluates to 0 if not.  

A suitable G\"odel numbering for the foregoing Boolean expressions can be formulated as follows.  Assign a symbol number to each of the symbols according to ${\it SN}(\lnot)=1$,  ${\it SN}(\lor)=2$, ${\it SN}(\land)=3$, ${\it SN}(\ (\ )=4$, ${\it SN}(\ )\ )=5$, and, for each $i\ge 1$, ${\it SN}({\bf e}_i)=5+i$. Then, where Boolean expression $E$ is the sequence of symbols $s_1,\ldots,s_n$, set the G\"odel number for $E$ as \medskip

${\it GN}(E) = \prod_{i=1}^n p_i^{SN(s_i)}$, \medskip

\noindent where $p_1, p_2, \ldots$ is the sequence of prime numbers $2,3,5,\ldots$. 

Considering this, we now turn our attention to the particular case of a Turing machine that computes the function$S$.  The forgoing definition of the language {\it SAT} was adopted from the textbook by Sipser \cite{sipser:2006}.  Siper's definition on p. 271 is: \medskip

${\it SAT}=\{\langle\phi\rangle|\phi {\rm\ is\ a\ satisfiable\ Boolean\ expression}\}$, \medskip

\noindent where $\langle\phi\rangle$ represents an encoding of the expression $\phi$ as a string of symbols.  This idea of encodings in introduced in \cite{sipser:2006} on p. 157.  Here, however, Boolean expressions already are strings of symbols, so this notion of encoding is not needed.  Thus, we are justified in using the foregoing definition (taking $E$ in place of $\phi$). The problem of deciding whether a given Boolean expression is in the language {\it SAT} is also referred to by the name {\it SAT}.  It is well-known that {\it SAT} is in NP, the class of nondeterministic polynomial time problems. This in fact is consequence of the version of the Cook-Levin Theorem that Sipser states on p. 276, namely, ``Theorem 7.37. {\it SAT} is NP-complete."  On p. 266, Sipser proves \medskip

\begin{quote}
{\bf Theorem 7.20.} A language is in NP iff it is decided by some nondeterministic polynomial time Turing machine.  \medskip
\end{quote}

\noindent Here it is implicit in the proof that this refers to a ``single-tape'' Turing machine.  

What is meant by the {\it time} of a Turing machine is given on p. 248 of \cite{sipser:2006} as {\bf Definition 7.1}: 

\begin{quote}
Let $M$ be a deterministic Turing machine that halts on all inputs.  The {\it running time} or {\it time complexity} of $M$ is the function $f:\mathcal{N}\to\mathcal{N}$, where $f(n)$ is the maximum number of steps that $M$ uses on any input of length $n$.  If $f(n)$ is the running time of $M$, we say that $M$ runs in time $f(n)$ and that $M$ is an $f(n)$ time Turing machine.  Customarily we use $n$ to represent the length of the input.  
\end{quote} 

\noindent Given this we can next invoke the following, stated on Sipser's p. 256:  \medskip

\begin{quote}
{\bf Theorem 7.11.}  Let $t(n)$ be a function, where $t(n)>n$.  Then every $t(n)$ time nondeterministic single-tape Turing machine has an equivalent $2^{O(t(n))}$ time deterministic single-tape Turing machine. \medskip
\end{quote}

The definition of the $O$ notation is given on p. 249 of Sipser as {\bf Definition 7.2}.  This reads as follows:   

\begin{quote}
Let $f$ and $g$ be functions $F,G:\mathcal{N}\to\mathcal{R}^+$. Say that $\bf f(n)=O(g(n))$ if positive integers $c$ and $n_0$ exist such that for every  $n\ge n_0$ \medskip

$f(n)\le c\ g(n)$. \medskip

When $f(n)=O(g(n))$ we say that $g(n)$ is an {\it upper bound} for $f(n)$, or more precisely that $g(n)$ is an {\it asymptotic upper bound} for $f(n)$, to emphasize that we are suppressing constant factors.
\end{quote}
       
\noindent The statement $ f(n)=O(g(n))$ is also often expressed by $f(n)\in O(g(n))$, expressing that $f(n)$ belongs to the class of functions that are in this sense bounded by $g(n)$, e.g., see the textbook \cite{cormen:1990}, p. 26. 

It follows that ${\it SAT}\in O^{g(n)}$ for some polynomial $g(n)$, where $n$ is the size of the input, i.e., the length of (number of symbols in) the Boolean expression whose membership in SAT is being decided.  This $g(n)$ employs only the usual operations of arithmetic---addition, subtraction, multiplication, division, and exponentiation to an integral or fractional power---all of which are primitive recursive, and thus clearly is primitive recursive.  Thus we have a primitive recursive bound on the number of steps that the Turing machine must take in order to make this determination.

However, we must also take into consideration that our function $S$ as defined in the foregoing additionally employs a front-end, so to speak, that decodes the G\"odel number of the Boolean expression before sending this expression to the procedure that carries out the decision.  So we must equip our Turing machine that decides {\it SAT} with a front end that performs this same operation.  It should be clear that, whatever is the time complexity of this front-end procedure, it will be another function of the aforementioned arithmetic operations, and therefore also primitive recursive.  Let us call this $h(n)$.  Then the sum $g(n)+h(n)$ serves as a primitive recursive bound on the number of steps that the Turing machine we have described requires to complete its execution.  This proves: \medskip

{\bf Theorem 3.}  The function SAT is primitive recursive. 

\section{The Class NP}

NP denotes the class of problems that can be solved in polynomial time on a non-deterministic Turing machine. This is Siper's Theorem 7.20, p. 266, where a problem is represented as the decision problem for some language, the words of which are symbolic strings that encode the objects of concern.  For example, the problem {\it HAMPATH} is that of finding a path in a directed graph from a given start node to a given end node and that goes through each graph node exactly once.   Such a path is called a {\it Hamiltonian path}.  This problem is given in \cite{sipser:2006}, p. 264, as the language \medskip

${\it HAMPATH}=\{\langle G,s,t \rangle|G{\rm \ is\ a\ directed\ graph\ with\ a\ Hamiltonian\ path\ from\ }s$ \hfill\break ${\rm \ to\ }t\}$. \medskip

The problem can be depicted as a function of natural numbers in the same manner as just described for {\it SAT}.  Symbol numbers are assigned to the symbols employed by the graph encodings, enabling G\"odel numbers to be assigned to the objects of concern (directed graphs with start and end points).  Then, similarly as for {\it SAT}, a function $H$ can be defined that takes such a G\"odel number as input and determines whether it is the number of a graph with the indicated Hamiltonian path, and this function $H$ can be shown to be primitive recursive in the same manner as in the foregoing.  Clearly, each problem in $NP$ has such a corresponding function of natural numbers. Thus, we have: \medskip

{\bf Theorem 4.} For any problem in {\it NP}, the corresponding function of natural numbers is primitive recursive. 

\section{Concluding Remarks}

None of the theorems proven in this paper should be surprising, as all are what one would intuitively expect.  Theorem 1 only seems reasonable.  So does its converse, Theorem 2.  Similarly, if one accepts Theorem 1, then Theorem 3 also seems natural, since the brute force method of resolving SAT amounts to computing the truth table, and there is no reason to think that an unbounded minimization operator should be required.  Theorem 4 is a natural extension of Theorem 3.  Accordingly, this paper has merely confirmed what many people that are knowledgeable regarding these topics have presumed to be true, despite the paucity or absence of published proofs.






\end{document}